Reduction of Spikes on the Sides of Patterned Thin Films for Magnetic Tunnel Junction Based Molecular Device Fabrication

Pawan Tyagi[1*], Edward Friebe[1], Beachrhell Jacques[1], Tobias Goulet[1], Stanley Travers[1]

[1]*University of the District of Columbia, Department of Mechanical Engineering, 4200 Connecticut Avenue NW Washington DC-20008, USA*

**ABSTRACT**

Sputter thin film deposition after photolithography often produces unwanted spikes along the side edges. These spikes are a significant issue for the development of magnetic tunnel junction (MTJ)-based memory and molecular spintronics devices, microelectronics, and micro-electro- mechanical systems because they influence the properties of the other films deposited on the top. Our molecular spintronics devices that utilizes MTJ as the testbed are almost short-lived and encountered high background current that masked the effect of molecular transport channels placed along the sides of MTJs. Therefore, tapered thin film edges are critically needed in devices. Here, we report a very cost efficient and fast way of creating an optimum photoresist profile for the production of 'spike-free' patterned films. This approach is based on performing a soaking in the photoresist developer after baking and before the UV exposure. However, the success of this method depends on multiple factors accounted for during photolithography - photoresist thickness (spin speed), baking temperature, soaking time and exposure time. Our recent experiments systematically studied the effect of these factors by following the L9 experimental scheme of the Taguchi Design of experiment (TDOE). The L9 experimental scheme effectively accommodated the study of four photolithography factors, each with three levels. After performing photolithography as per L9 TDOE, we conducted sputtering thin film deposition of 20 nm Tantalum. Then we conducted an atomic force microscope (AFM) study of thin film patterns and measured the spikes along the edges of the deposited Tantalum. We utilized spike height as the desired property and chose "smaller the better" criteria for TDOE analysis. TDOE enabled us to understand the relative importance of the parameters, relationship amongst the parameters, and impact of the various levels of the parameters on the edge profile of the thin film patterns. We discovered that baking temperature was the most influential parameter; presoak time and photoresist thickness were two other influential factors; exposure time was the least effective factor. We also found that 4000 rpm, 100 C soft baking, 60 s soaking and 15 s UV exposure yielded the best results. Finally, the paper also discusses the interdependence of selected factors, and impact of the individual levels of each factor. This study is expected to benefit MEMS and micro/nanoelectronics device researchers because it attempts at finding a cheaper and faster alternative to creating an optimum photoresist profile.

**INTRODUCTION**

Liftoff is a critical step for the development of advancing microelectromechanical systems [1, 2], nanoscale material fabrication [3], and tunnel junction based molecular devices [4]. Liftoff is extremely useful for the materials which are challenging to pattern by conventional lithography and etching techniques. In general, the lift-off approach is an economical and much simpler method to pattern thin films for different applications. Liftoff is a process to remove unwanted photoresist to free up a thinfilm deposited in a photoresist cavity. During liftoff, the photoresist layer under the film is removed with a solvent,. This step leaves the film which was deposited directly on the substrate into the phoresist cavity. So far liftoff has been performed in three ways: the single-layer method [5], the multi-layer method [6] and the surface-modified method [7, 8]. The single-layer method, involve a negative tone photoresist. This single layer based liftoff is the simplest and involves only one lithography step. However, this approach has several demerits: (i) high cost of utilizing negative photoresist is costly, (ii) the delicate process conditions of the negative photoresist, and (iii) the difficulty in the removal of the negative photoresist resulting in a rough sidewall profile of the metal. The multi-layer method make microfabrication costly and relatively complex to perform. This multi-layer approach requires the deposition of a separate undercut producing chemical under the main photoresist. The chemical under photoresist dissolves in developer faster than the exposed photoresist. Hence, an undercut profile get generated. Several forms of multilayer based approaches has been published to give smooth lift off [9, 10]. In the case of the surface-modified method, the top surface of the photoresist is chemically modified by soaking it in chlorobenzene, toluene or tetramethyl ammonium hydroxide (TMAH) solution to develop it at a slower rate than the underlying resist. These conditions are known to be somewhat tricky, and there are potential health hazards when using these solvents [10]. Recently, a simpler approach was published where a glass diffuser was utilized to develop undercut profile in the photoresist [10]. However, it is necessary to utilize a glass diffuser as an additional component. In this paper, we discuss a simple and effective approach to perform smooth lift off to produce patterned thin films without creating spikes/notches along the side edges of metal films.

Our approach reported in this paper only requires an additional soaking step of soft baked photoresist in the developer solution for ~1 min. It is well know that basic photolithography steps are the following: (i) spin coating of positive tone photoresist on a flat substrate, (ii) soft-baking of the photoresist, (iii) UV exposure of the photoresist through mask, and finally (iv) developing in a developer chemical. Our approach is based on conducting an additional pre exposure soaking step in the developer solution. Hence we call this step pre-soak step now onwards. This step will harden the top surface of the photoresist. As a result during the final developing step, after the exposure, pre-soak affected top surface of the photoresist will dissolve slower than the photoresist away from the top surface. This differential developing rate of the pre-soak affected area and photoresist away from the surface creates an undercut profile. During directional thin film deposition in the photoresist cavity with an undercut profile material will not touch the photoresist sidewall to produce notch or spike on the edges of patterned thin films. However, finding an optimum set of parameters for developing undercut profile required the process optimization. We have applied Taguchi design of experiment (TDOE) to find the most promising experimental conditions and also to gain the deeper insight about the impact of process parameters. Here we also demonstrate the utility of adopting the pre-soak based photolithography for realizing tunnel junctions with a ~2 nm tunnel barriers.

**EXPERIMENTAL DETAILS**

For this study we utilized an oxidized silicon wafer and performed photolithography with an aim of creating undercut profile. We utilized Shipley 1813 positive tone photoresist and MF319 developer solution for this study. Exposure step was performed with Karl Suss MJB3 mask aligner. To produce undercut profile we varied key photolithography parameters such as spin coating speed, baking temperature, pre-soaking duration in the developer, and exposure time. Each photolithography step possessed three levels each as indicated in the Table 1.

**Table1.** Levels assigned to the parameters of interest for the Taguchi Design of Experiment

| Parameter | Level 1 | Level 2 | Level3 |
|---|---|---|---|
| Spin Coating Speed (rpm) | 2000 | 3000 | 4000 |
| Soft Baking Temperature ($^0C$) | 80 | 90 | 100 |
| Developer Soaking Time (sec) | 40 | 60 | 80 |
| UV Exposure Time (sec) | 15 | 20 | 25 |

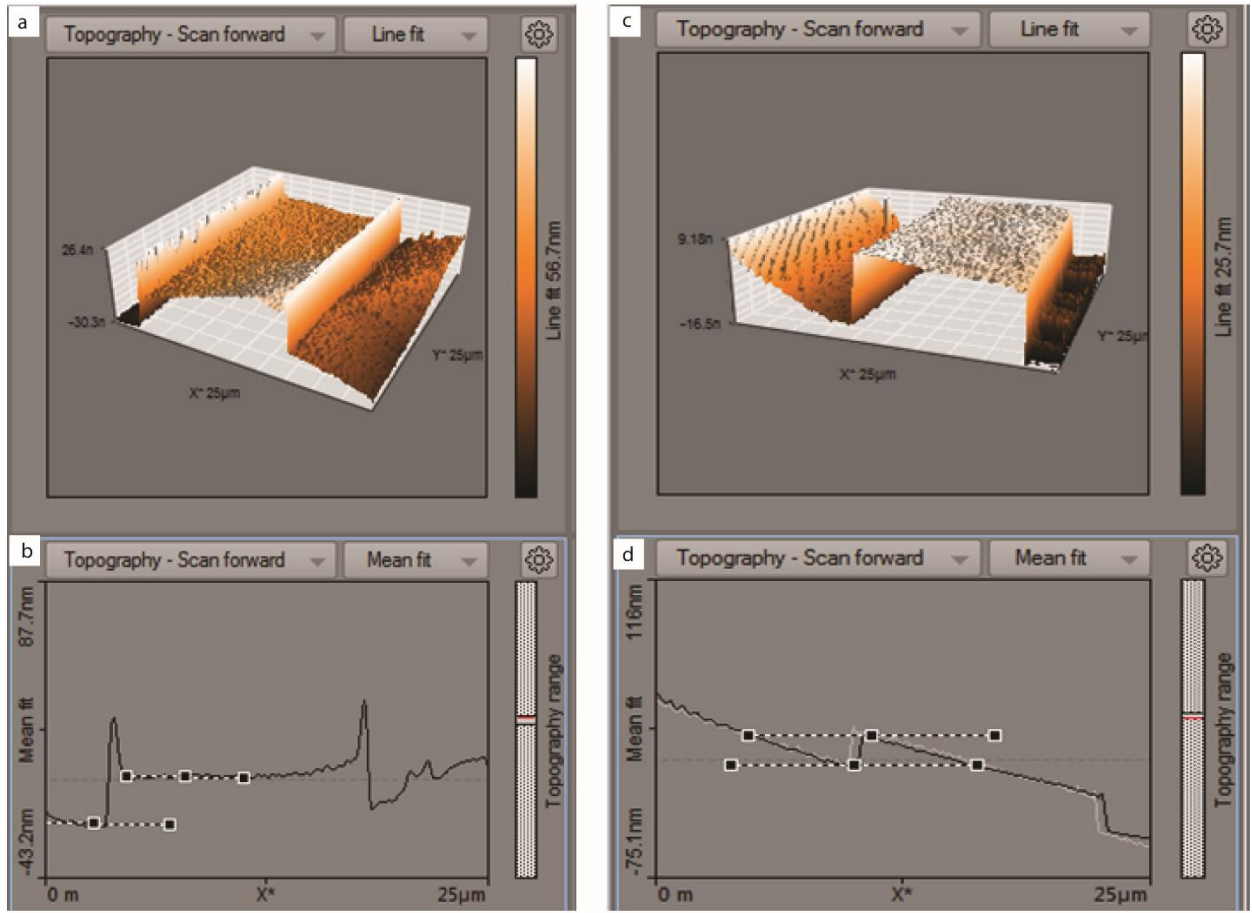

**Figure 1** AFM images notches before and after the optimization of photolithography process. (a) 3D image and (b) cross section view of patterned thin film with notches along the side edges. Side edges are result of improper photolithography conditions (c) 3D and (cross sectional view of patterned thin film with no sharp notches along the side edges. Notch free edges are result of optimized photolithography parameters.

After each photolithography experiment ~20 nm thick tantalum film was deposited in the photoresist cavity. For the thin film deposition AJA International Orion 5 sputtering system was utilized. Thin film was performed with a RF powered gun at 2 mtorr argon pressure and 100 W gun power. After thin film deposition liftoff was conducted in Shipley resist remover and atomic force microscopy (AFM) was performed to measure the step-edge profile. For this study Naio-AFM was utilized. For every experiment average notch thickness on the patterned film edge was recorded. Based on notch height data we conducted TDOE data analysis by utilizing smaller the better criteria. The results of analysis are discussed in the following section.

**RESULTS AND DISCUSSIONS**

We conducted TDOE data analysis by utilizing smaller the better criteria because our intention is to minimize the height of notches along the patterned thin film edges. TDOE provided an optimum comobinations of levels for the parameters listed in Table 1. Photolithography performed with the optimized parameters enabled the realization of patterned thin film with virtually no side notches. Figure 3 shows the AFM images of the patterned thin film photolithography process without and without a pre-soaking step in developer solution. Thin film deposited in the photoresist cavity prepared without a pre-soaking step always produced notches along the side edges (Fig. 1 a and b). The side notches can be as tall as few hundred nm. When a ~ 2 nm tunnel barrier is deposited on the top of bottom electrode then tall notches pierce through tunnel barrier and touch the top metal electrode to create short circuit. After including a pre-soaking step in the developer solution right before exposure we achieved notch free bottom electrodes. The AFM images showed the ~20 nm thick Ta film deposited in the photoresist cavity, prepared with the photolithography process with a pre-soaking step, yielded notch free patterned films. The tunnel junctions prepared on the bottom electrode without notches produce high quality tunneling behavior. However, to find the best combination of levels for different photolithography parameter we utilized Taguchi design of experiment (TDOE). This approach is very effective in reducing the number of experiments required to gain insight. For the present case where four variables are to be optimized L9 TDOE was found to be most suitable [Ref]. We utilized Qualitek-4 software to design the optimization study. According to TDOE scheme nine experiments with the following levels were obtained for process optimization (Table-2).

**Table 2**. *L9 experiments for the bottom layer optimization*

| Experiment Number | Spin Speed (rpm) | Baking Temperature ($^0$C) | Soaking Time (sec) | UV Exposure Time (sec) | Taguchi analysis data | S/N ratio Based on AFM data |
|---|---|---|---|---|---|---|
| 1 | 2000 | 80 | 40 | 15 | $y_1$ | -38.136 |
| 2 | 2000 | 90 | 60 | 20 | $y_2$ | -38.517 |
| 3 | 2000 | 100 | 80 | 25 | $y_3$ | -38.289 |
| 4 | 3000 | 80 | 60 | 25 | $y_4$ | -19.905 |
| 5 | 3000 | 90 | 80 | 15 | $y_5$ | -37.912 |
| 6 | 3000 | 100 | 40 | 20 | $y_6$ | -22.304 |
| 7 | 4000 | 80 | 80 | 20 | $y_7$ | -33.086 |
| 8 | 4000 | 90 | 40 | 25 | $y_8$ | -36.51 |

| 9 | 4000 | 100 | 60 | 15 | y₉ | -10.59 |

Using the experimental data for each run (Table 2) the data analysis was performed. Initially, we computed the effect of different levels of four parameters (Fig. 3). We calculated the level total and their averages. To acomplish this objetive, results of all trials involving the particular level of a parameters are added, and then dividing by the number of data points added. For instance, for the three levels of spin speed paramer the following equations were adopted [11].

$A1$ = Spin speed (2000) = $y_1+y_2+y_3$      (1)
$A2$ = Spin speed (3000) = $y_4+y_5+y_6$      (2)
$A3$ = Spin speed (4000) = $y_7+y_8+y_9$      (3)

Subsequently, the averages $\bar{A}(1)$, $\bar{A}(2)$, and $\bar{A}(3)$ were calculated by dividing A(1), A(2), and A(3) by three, since each level of time parameter appeared in the three trials (Table 2). The values of $\bar{A}(1)$, $\bar{A}(2)$, and $\bar{A}(3)$ were -37.28, -32.64, and -28.54, respectively. To evaluate the impact of each level a referece data was obtained by averaging $\bar{A}(1)$, $\bar{A}(2)$, and $\bar{A}(3)$. Likewise average values of each level of all the parameters were calculated. The average values of each level of all the parameters were plotted to provide the insight about their relative impact (Fig. 2). After the TDOE analysis we plotted data for each factor (Fig. 2). The 4000 rpm spin speed turned out to be the most influenitial (Fig. 2a). On the otherhand spin speed 2000 was least effective in providing notch/spike free thin film edges. The photoresist baking temperature parameter showed interesting non linear pattern. TDOE analysis suggested that 100 $^0$C baking temperature was the most impactful in ealizing notch/spike free edges (Fig. 2b). On the otherhand 90 $^0$C was least effective as compared to other two levels (Fig. 2b). The pre-soak time factor also showed non linear dependence on the three levels (Fig. 2c). Soaking of soft baked photoresist in developer solution for 60 sec yielded the highest impact. Other two soak periods produced comparable effect and low impact intensity. The UV-exposure time produced relatively low impact intensity for all the three levels. However, 15 second exposure time produced the heighest impact intensity as compared to the other two levels (Fig. 2d).

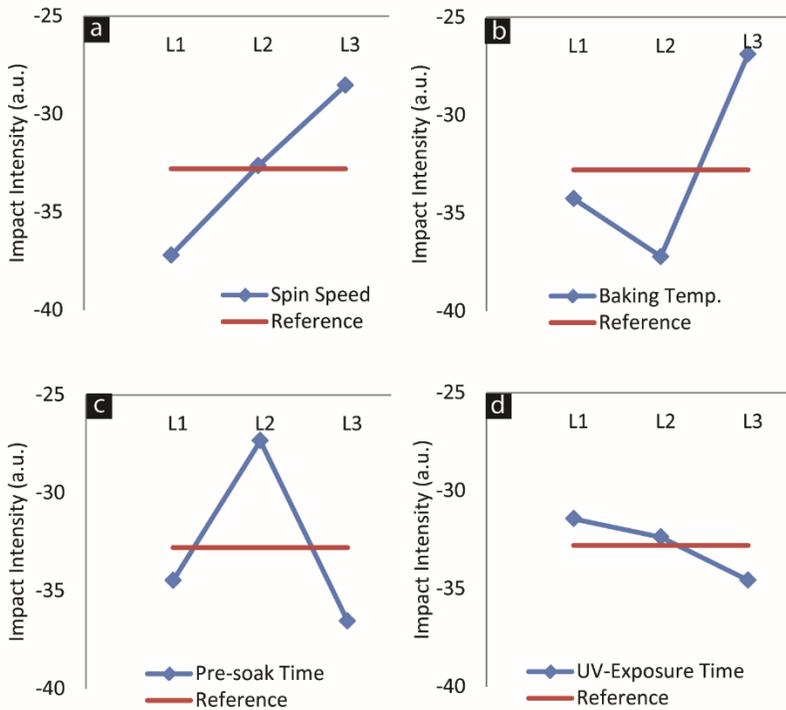

Fig. 2 Effect of different levels of factor (a) spin speed, (b) baking temperature, (c) pre-soak time in developer, and (d) UV exposure time.

It is noteworthy that Taguchi analysis is also capable of dertermining the degree of interaction among factors [12]. We made six pairs ($^4C_2$) by selecting two factors at a time using permutation-combination rule. Taguchi analysis rank the degree of interaction in terms of % interaction severity index [12]. The pre-soak-time and UV-exposure time appeared had the strongest interaction and exhibited 65.47 SI (Table 3). High SI suggests that any variation in any one of the factor will strongly impact other factor. The spin speed and baking temperature exhibited the second highest SI with a magnitude of 25.97 (Table 3). It is noteworthy that there is 40 SI between the top two interactions. The interaction between baking temperature and exposure time factors was the waekest with a SI of 4.18 only (Table 3). It means varying anyone of the baking temperature and UV exposure will not impact each other strongly. It is noteworthy that these interpretations are valid for the range of different levels utilized in this study.

**Table 3:** *Degree of interaction between two factors is determined by the percent severity index (SI).*

| Interacting variables | SI(%) |
|---|---|
| **Pre-soak time x UV exposure time** | 65.47 |
| **Spin speed x Baking Temp.** | 25.97 |
| **Spin speed x UV exposure time** | 17.38 |
| **Baking Temp. x Pre-soak time** | 16.73 |
| **Spin x Pre-soak time** | 12.04 |
| **Baking Temp. x UV Exposure Time** | 4.18 |

In order to calculate the relative impact of the four parameters we performed Analysis of variance (ANOVA) [12]. We calculated sum of all the the results and denoted it by T. the magnitude of *T* was found to be -275.49. Subsequently a correction factor (*C.F.*) was calculated by dividing the square of sum by the total number of experiments ( *C.F.* $=T^2/9$). We computed sum of squares (S) by using the following equation: $S=T^2-C.F.$ To proceesd to the next step sum of squares due to a parameter were calculated. For instance sum of square for the parameter spin speed (A) were calculated using equation (4).

$SS_A=[A(1)^2/3 + A(2)^2/3 + A(3)^2/3]-C.F.$      (4)

Similarly, the sum of squares for three other parameters soft baking temperature (B), pre-soak time (C), and UV exposure time (D) was computed. In the next step error sum of squares ($SS_e$) was calculated by utilizing the equation (5).

$SSe=SS_T - (SS_A+SS_B+SS_C+SS_D)$      (5)

The degree of freedom for all the parameters were calculated. Total degree of freedom was eight; for each parameter degree of freedom was two (number of levels-1). Subsequently, mean square of variance was clculated for different parameters by dividing the sum of squares with the degree of freedom for the individula parameters. The mean square of variance for the error was calculated by dividing the *SSe* with degree of freedom for the error. Nextly, we computed pure sum of square by following the following equation (6)

$$SS'_A = SS_A - (V_e \cdot F_A) \quad (6)$$

Similarly, pure sums were calculated for other parameters too (Table 3). Finally contribution (*P*) for the individual parameters was calculated by utilzing the following equation.

$$P_A = SS'_A / SS_T. \quad (7)$$

Similarly, impact of other three parameters were computed and tablulated in the Table 3.

**Table 4** *ANOVA analysis of parametrs. The last column of the table shows the impact of individual parameter.*

| Factors | DOF(f) | Sum of Squares | Variance (V) | Pure Sum(S') | P(%) |
|---|---|---|---|---|---|
| Spin speed | 2 | 112.86 | 56.43 | 112.86 | 25.77 |
| Baking Temp. | 2 | 169.82 | 84.91 | 169.82 | 38.77 |
| Pre-soak time | 2 | 139.79 | 69.89 | 139.79 | 31.92 |
| UV Exposure time | 2 | 15.51 | 7.76 | 15.51 | 3.54 |

Taguchi analysis suggested that photoresist baking temperature was the most influential parameter for realizing notch free patterned thin films from a photoresist cavity. The baking temperature parameter accounted for a contribution of ~38.77%. The pre-soak time and spin speed were at the second and third place, respectively (Table 4). UV exposure time duration was found to be least influential. This study suggest that major focus of parameter optimization can be on three parameters.

Finally, we utilized Taguchi analysis to yield optimum combination of levels for four parameters to yield the optimum patterned film edge profile. According to the Taguchi analysis optimum combination of the factors included 4000 rpm spin speed, 100 $^0$ baking temperature, 60 seconds pre-soak time, and 15 sec UV exposure.

To double validate the TDOE results we conducted photolithography by utilizing the optimum levels for each factor. We indeed obtained the notch free patterned thin films. The 3D image and the cross sectional vie of the patterned thin film did not show noticeable spike or notches along the edges (Fig 1c and d). As mentioned earlier our ultimate objective was to form a metal-insulator-metal tunnel junction with ~2 nm thick insulator. We produced a tunnel junction sample by following the method described elsewhere[4, 13]. The optimized photolithography conditions were only applied to the bottom electrode (Fig. 3a). The thickness of top and bottom electrode was ~15 nm. The tunnel barrier was made up of ~ 2.5 nm thick alumina. The overall junction area was 50 µm$^2$. The current-voltage study on the tunnel junctions showed high quality tunneling behavior as shown in Fig. 3b.

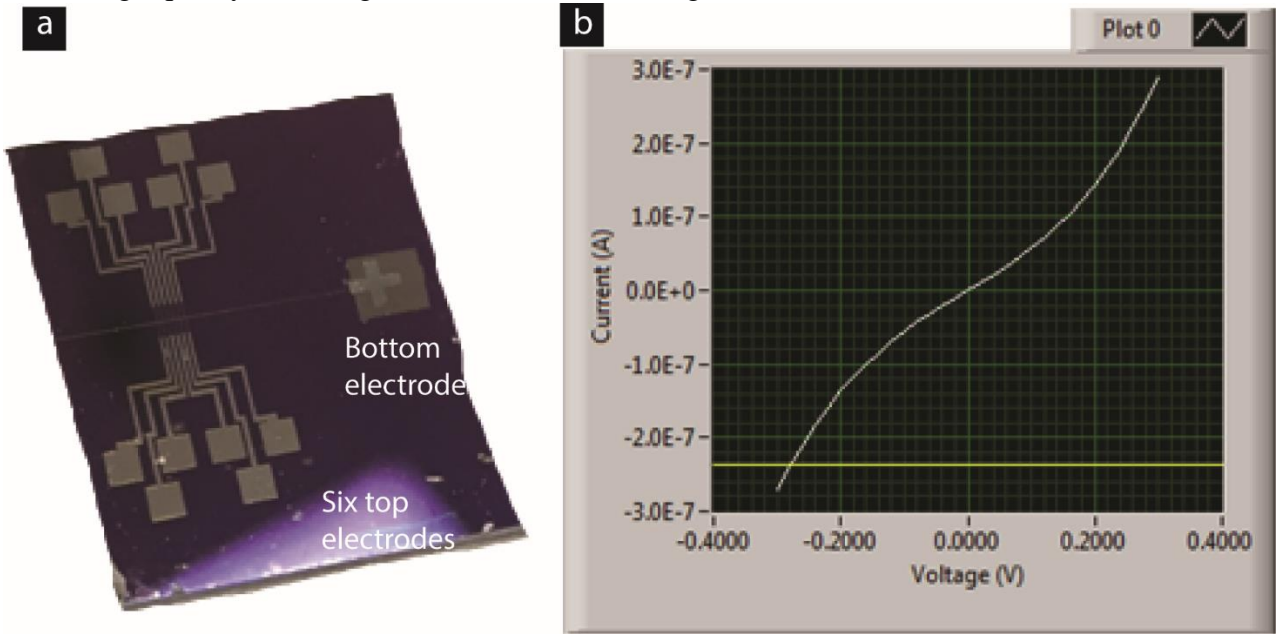

**Figure. 3** (a) Tunnel junctions where bottom electrode is deposited in the optimized photoresist protocol to remove notches along the edges. (b)Current-voltage study showing the realization of tunneling behavior between top and bottom metal electrodes.

## CONCLUSIONS

We have explored a simplified photolithography process to avoid the notches or spikes on the thin film edges. Our primary motivation is to produce a smooth edged bottom electrode for the metal-insulator-metal (tunnel junction) based devices [MTJMSD]. In these devices top and bottom metal electrodes cross each other at 90 degree and an ultrathin insulator is sandwiched between the two metal electrodes to form metal-insulator-metal. One of the most convenient route to pattern bottom electrode is to utilize photolithography approach. However, to form After the bottom metal electrode deposition an ~ 2 nm ultrathin insulator is deposited on the top of it. Avoiding notches on the bottom electrode is absolutely essential to avoid puncturing the tunnel barrier. A tunnel junction is made up of bottom metal electrode, a 2 nm thick tunnel barrier, and top metal electrode.

## ACKNOWLEDGEMENT

The study in this paper was in part supported by National Science Foundation-Research Initiation Award (Contract # HRD-1238802), Department of Energy/ National Nuclear Security Agency (Subaward No. 0007701-1000043016), and Air Force Office of Sponsored Research (Award #FA9550-13-1-0152). Pawan Tyagi thanks Dr. Bruce Hinds and Department of

Chemical and Materials engineering at University of Kentucky for facilitating experimental work on MTJMSD during his PhD. We also acknowledge the support of UDC STEM center for supporting this work. We thankfully acknowledge the research facility support from Center of Nanoscience and Technology –NIST Gaithersburg. Any opinions, findings, and conclusions expressed in this material are those of the author(s) and do not necessarily reflect the views of any funding agency.